\documentclass[12pt,preprint]{aastex}
\usepackage{epsfig}
\usepackage{graphicx}
\usepackage[T1]{fontenc}
\usepackage{natbib}
\usepackage{fmtcount}
\usepackage{rotating}

\newcommand{\igr}{IGR J17091--3624}
\newcommand{\grs}{GRS 1915+105}

\begin{document}

\title{Why is IGR J17091--3624 so faint? 
Constraints on distance, mass, and spin 
from `phase-resolved' spectroscopy of the `heartbeat' oscillations}

\author{Anjali Rao{\thanks {Email: anjali@prl.res.in}}}
\author{S. V. Vadawale}
\affil{Physical Research Laboratory, Ahmedabad-380009, India}

\begin{abstract}

IGR J17091--3624 is a transient X-ray source and 
is believed to be a Galactic
black hole candidate. Recently, it has received a considerable attention 
due to the detection of peculiar variability patterns known as `heartbeats', 
which are quasi-periodic mini-outbursts repeated over timescales ranging 
between 5 and 70 s. So far, such variability patterns have been observed only in
GRS 1915+105 and these are classified as $\rho$- and $\nu$-variability classes.
Here, we present the results of `phase-resolved' spectroscopy of the `heartbeat'
oscillations  of \igr{} using data from simultaneous observations 
made by RXTE and XMM-Newton. We find that the 0.7--35 keV spectra can be 
fitted with a `canonical' model for black hole sources consisting of only two
components---a multi-temperature disk black body and a power law (or its 
equivalent). We attempt to constrain the system parameters of the source by 
simultaneously fitting spectra during different phases of the burst profile 
while tying the system parameters across the phases. The results indicate that 
the source is a high inclination binary ($i$>53$^\circ$). Further, the observed
low flux from the source can be explained only if the black hole spin is very 
low, along with constraints on the black hole mass (<5 M$_\odot$) and the 
distance (>20 kpc). For higher inclination angles, which is favored by the 
data, the black hole spin is required to be negative. Thus low or retrograde 
spin could be the reason for the low luminosity of the source.

\end{abstract}

\keywords{accretion, accretion disks --- black hole physics 
--- X-rays: binaries --- X-rays: individual (IGR J17091-3624)}

\section{Introduction}
The micro-quasar \grs{} is an enigmatic black hole binary (BHB) exhibiting
enormous variability which have been classified in more than 14 different 
variability classes \citep{belloni,fenbel}. It is believed that the extreme 
variability and rapid state changes observed in \grs{} are due to a very 
high accretion rate, which is close to, or at times higher than, the 
Eddington accretion rate \citep{done}. It is also known for exhibiting 
large superluminal radio flares and steady radio emission which are always 
associated with specific X-ray variability classes 
\citep{mirabel,fender,vadawale}. 
Such an extreme and correlated multi-wavelength variability makes \grs{} 
a unique BHB.
In this context, \igr{}, a new X-ray transient source believed to 
be a BHB, generated considerable interest recently. It was detected by 
Integral/IBIS in 2003 \citep{kuulkers}. It has exhibited repeated outbursts 
with periods of two to four years in 1994, 1996, 2001, 2003, 2007, and 2011 
\citep{revnivtsev,kuulkers,capitanio2,capitanio3,krimm,krimmkennea}. 
The recent 2011 outburst of \igr{} was unusually long and the source was 
found to be active even after one year \citep{altamirano1}.
During this outburst, \igr{} revealed its highly variable nature and showed 
variability patterns so far observed only in \grs{}. The most prominent of 
these patterns was the `heartbeat' pattern, similar to the $\rho$-class in 
\grs{}. \citet{altamirano2}
documented the first six months of RXTE observations and showed that not 
only $\rho$-class but many other variability patterns similar to $\nu$-, 
$\alpha$-, $\lambda$-, $\beta$-, $\mu$-, and $\chi$- classes have been 
observed during this outburst of \igr{}. \citet{altamirano3} also detected 
a high frequency quasi-periodic oscillation (HFQPO) in this source with a frequency of 66 Hz, 
which is almost identical to the frequency of HFQPO in \grs{}.
Despite striking morphological similarities, the most perplexing difference 
between the two sources lies in their observed intensities. While \grs{} is
one of the brightest X-ray sources with a typical brightness of $\sim$0.5 -- 
2 Crab, \igr{} is about 20 times fainter. In the present scenario, mass, 
distance, and inclination for this source are rather poorly constrained, 
with reports so far suggesting a mass range of <3 M$_\odot$ \citep{altamirano2} 
to $\sim$15 M$_\odot$ \citep{altamirano3} and a distance range of $\sim$11 
kpc \citep{rodriguez} to $\sim$20 kpc \citep{pahari}. Nevertheless, the 
apparent faintness of \igr{} is difficult to explain even after assuming 
the smallest possible mass of 3 $M_\odot$ for a black hole \citep{fryer} 
and the largest possible distance of $\sim$25 kpc for a Galactic source.
Here, we attempt to investigate the possible reasons for this apparent 
faintness of \igr{} by simultaneously fitting spectra at different phases. 
The main idea is that the system parameters cannot change over the phase 
of the oscillations. Therefore, a simultaneous fitting of spectra at 
different phases, with system parameters tied across phases,  may put a 
better constraint on them. This, along with a proposal that the `heartbeats'
can be used as a `standard candle',  leads to a primary conclusion that
the faintness of \igr{} is due to its low or negative spin.

\section{Observations and Data Analysis}
We have used data from long simultaneous observations of \igr{} made on 
2011 March 27 with RXTE (ObsID: 96420-01-05-000, total exposure $\sim$21 ks) 
and XMM-Newton (ObsID: 0677980201, total exposure $\sim$39 ks) with net
simultaneous exposure of $\sim$15 ks.
The data reduction for the RXTE/PCA observation was carried out with HEASoft 
version 6.8 following standard analysis procedure for Good Xenon data. 
We extracted 1 s light curve from PCU2 data. It showed the typical 
$\nu$-class oscillations with periods ranging from 30 to 50 s (Figure 1). 
It contained a total of 385 bursts.
We carried out `phase-resolved' spectroscopy for these bursts in the 
energy range of 3.0--35.0 keV for RXTE/PCA and 0.7--12.0 keV for XMM/PN 
data as described below.
The peak time for 
each burst was identified in a semiautomatic manner using an IDL 
script and the peak-to-peak interval between consecutive bursts was 
divided into 64 phases of equal length. The start and stop times of each
phase, recorded in RXTE mission time for 385 bursts, were used 
for extracting spectra for each phase. Total counts for all 64 
spectra and their corresponding exposure times were then used to generate 
the `phase-folded' light curve (Figure 2). 
The 64 phase bins were grouped into five phases as shown in Figure 2 and 
the spectra extracted for
these five phases were used for simultaneous spectral fitting. The grouping
was carried out mainly by the visual inspection of the folded RXTE/PCA 
lightcurve.
The XMM observation was carried out in the {\em fast timing} mode of 
EPIC-MOS and the {\em burst} mode of EPIC-PN and we followed the standard 
analysis procedures for these modes using {\em SAS v11.0.0} and the latest
calibration files.
We used data from XMM-PN only because MOS2 data could not be checked for 
possible pileup (generation of pattern plot always resulted in error) whereas
MOS1 data are not useful in timing mode because of a dead pixel in the CCD.
For PN data, the observed and the expected pattern behavior differed below 
0.7 keV and hence the energy range for rest of the analysis was restricted to 
0.7--12.0 keV. 
Start and stop times of the 64 phases of all bursts from RXTE mission were 
converted into XMM mission time using the {\em xTime} tool, available at 
HEASARC, which were used to build gti files using SAS task 
{\em gtibuild}. These gti files were used for extracting the 64 phase spectra 
using the task {\em evselect}. The `phase-folded' light curve was generated
using the total counts and the exposure times, as described earlier. The subtle features were averaged out
as a consequence of the quasi-periodicity and co-adding, but the 
overall profile of  
the `phase-folded' light curve followed a typical burst cycle. 
Further, it was 
seen that the oscillations were more pronounced in the XMM light curve 
indicating that the accretion disk radiation was primarily
participating in the oscillations and not the Comptonized emission from
the corona which dominates at higher energies. 
Source spectra from the five grouped phases were extracted using RAWX columns 
between 32 and 42 for single and double pixel events (pattern $\leq$4). 
SAS tools {\em rmfgen} and {\em arfgen} were used to generate redistribution 
matrices and ancillary files, respectively, and
the same files were used for the spectra of all phases. 
The background spectrum was extracted from RAWX columns between 5 and 7 after
confirming that the region was not contaminated significantly with source 
photons in the selected energy range. A single background spectrum was used 
for the five phases. All spectra were rebinned using sastask {\em specgroup} 
to have a minimum of 25 counts per channel. 

\subsection{Simultaneous spectral fitting}

Once we extracted the spectra for the five phases for both PCA and PN, the 10 
spectra were fitted simultaneously with various parameters tied as follows. The
10 spectra were loaded into XSPEC as 10 data groups. For a given phase, all 
parameters for PCA and PN spectra were tied together except the normalization 
constant which was frozen at 1.0 for PN whereas for PCA it was kept free but 
tied across the five phases. For a particular spectral model, we tied all 
parameters representing system property, such as mass, distance, inclination, 
spin or combination of these, across the five phases. The parameters describing
the accretion process such as inner disk temperature or accretion rate, inner
disk radius, were fitted independently for each phase.

%begin{center}
\begin{figure*}
\centering
    \includegraphics[width=130mm,height=80mm]{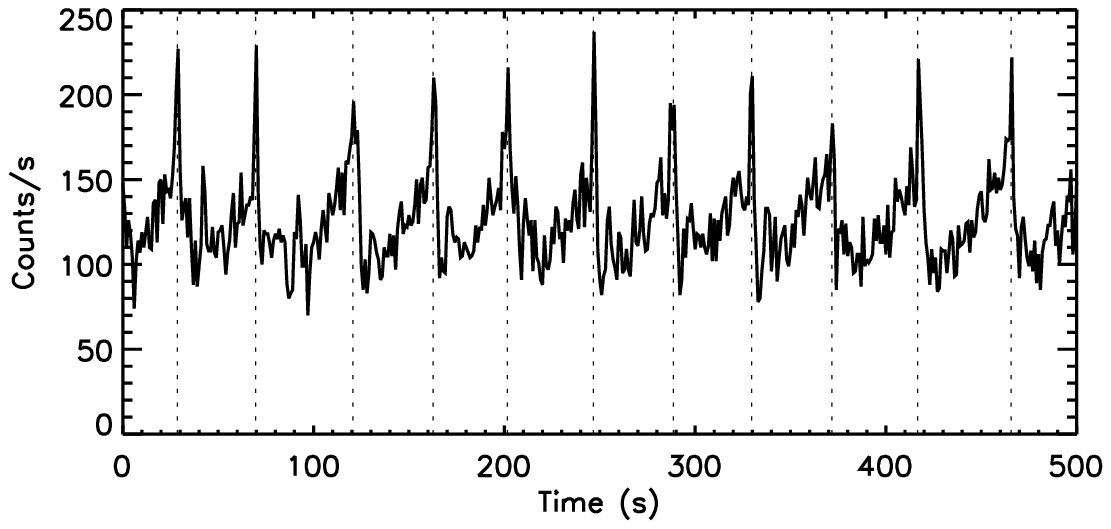} 
    \caption{RXTE/PCA light curve of IGR J17091-3624 from observation 
    96420-01-05-000. The light curve was extracted with a bin size of 1 s from 
	PCU2 covering an energy range of 2.0--60.0 keV. The vertical lines show
the identified peak times of bursts.}
\end{figure*}
%end{center}

\section{Results and Discussion}

Wide-band X-ray spectrum of a black hole binary generally consists of two 
dominant components: a multi-color disk and a high-energy tail arising from 
Compton scattering in an optically thin region surrounding the disk. 
We used {\tt DISKPN} \citep{gierlinski} as a simplified disk model, 
primarily because its 
parameters are cleanly separated in accretion-process-dependent 
parameters (disk temperature and inner disk radius) and system
parameters (normalization). 
For a general relativistic description of 
the multi-temperature disk spectrum, we used {\tt KERRBB} \citep{li}, 
which is widely used for its accurate modeling of disk 
spectrum and to investigate black hole spin.
The high-energy tail of the spectrum is typically modeled as 
{\tt POWERLAW} to approximate the Comptonized component. However, 
\citet{steiner} proposed a more physical model {\tt SIMPL} 
to empirically describe the Comptonized component. Here, we have used
{\tt SIMPL} along with one of the two disk models
({\tt DISKPN} and {\tt KERRBB}) to fit the spectra in the five phases 
simultaneously, as described below.

\begin{figure}
\centering
\includegraphics[width=110mm,height=60mm]{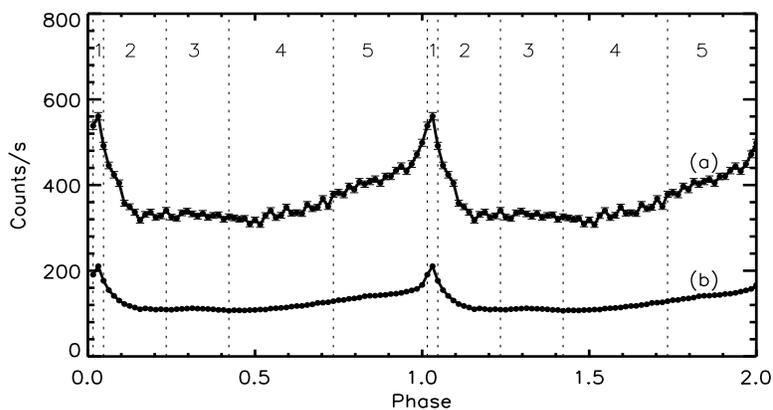} 
\caption{Phase-folded light curves from  XMM-Newton and RXTE,
labeled (a) and (b), respectively.
Dotted vertical lines show demarcations for the five phases used 
for the simultaneous spectral fitting.}
\end{figure}

\subsection{Model: {\tt DISKPN}}
For the first part of the analysis, we fitted RXTE-PCA and XMM-PN spectra for
the five phases simultaneously with {\tt CONST*PHABS*(SIMPL$\otimes$DISKPN)}. 
The parameter CONST was used to account for possible 
calibration uncertainties  between the two instruments. Normalizations
of {\tt DISKPN} were tied across the five phases, whereas rest of the 
parameters were allowed to vary independently. Though the interstellar 
absorption can be considered as a part of system parameters, we allowed it 
to vary to account for any phase-dependent absorption intrinsic to the 
source. We, however, found that the N$_H$ values for the five phases were not
significantly different and the fitted values of N$_H$ were in agreement with 
the values reported by \citet{rodriguez} and \citet{krimm}. Table 1 provides 
the results of spectral fits. It can be seen that the inner disk temperature 
is highest for phase 1 corresponding to the peak of the bursts, implying
higher accretion rate as expected. We verified that neither the Fe-line nor 
the reflection component was required to fit the data. A best fit was 
obtained with $\chi^2$ value of 1709.0 for 1030 degrees of freedom. 
However, for the present work, more important is the best-fit value of 
{\tt DISKPN} normalization, $N_{DPN}=\frac{M^2 \cos(i)}{D^{2}f_{col}^{4}}$,
as it can provide some constraints on mass ($M$), distance ($D$), 
and inclination ($i$).
The best fit value of {\tt DISKPN} normalization was found to be 
$N_{DPN}=4.0\times10^{-4}$.
Assuming a minimum mass of $\sim$3 M$_\odot$ and a maximum distance of 
$\sim$25 kpc for a Galactic black hole candidate along with a standard 
value for color correction factor, $f_{col} = 1.7$, the best-fit
value of $N_{DPN}$ resulted in a lower limit on inclination angle  
of 76$^\circ$. Considering the 90\% upper confidence limit
of $1.04\times10^{-3}$ for the normalization, the lowest 
possible inclination angle was $\sim$53$^\circ$. 
This lower limit comes only from simultaneous spectral fitting and it 
is not dependent on any additional information, and hence it can be considered
as fairly robust. Since spectral fitting could not constrain lower limit 
of N$_{DPN}$, it 
was not possible to obtain an upper limit on $i$ with spectral fitting.
However, simultaneous spectral analysis with {\tt DISKPN} to model 
accretion disk spectra suggests that \igr{} is a high
inclination binary. This is consistent with the finding of \citet{capitanio1} and \citet{king}.

\begin{figure}
\centering
 \includegraphics[width=120mm,height=120mm]{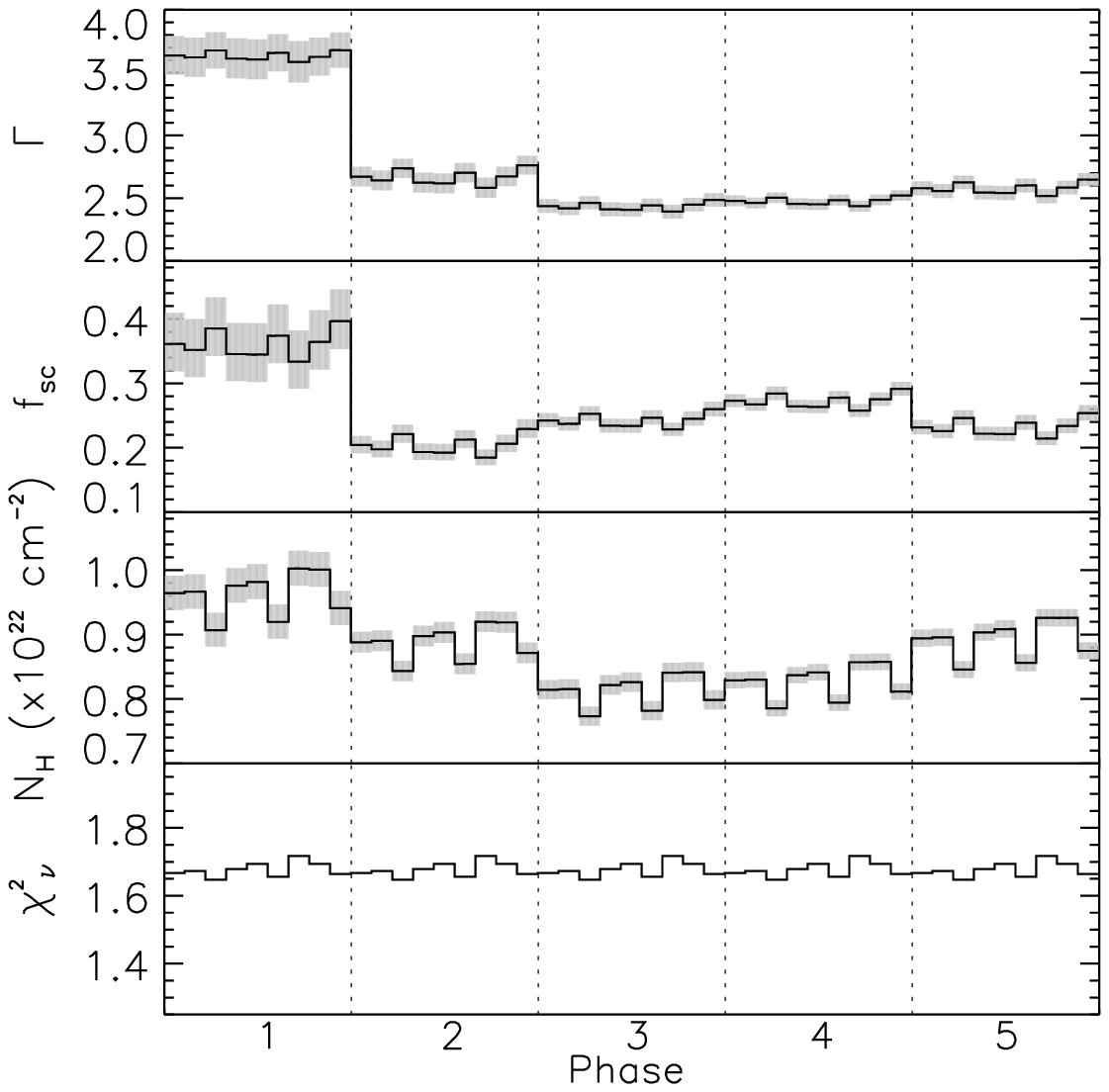} 
 \caption{Best-fit values for all five phases of $\Gamma$, scattering 
fraction (f$_{sc}$), N$_H$, and $\chi^2_\nu$ for all combinations of $D$, $M$, $i$ and 
$a_\ast$. The parameters are more or less
similar for any combination of system parameters and hence distinction
between them is not made. Error bars (90\% confidence) are shown with 
gray and only a subset of data are shown here.}
\label{allpar}
\end{figure}

\begin{figure}
\centering
 \includegraphics[width=120mm,height=120mm]{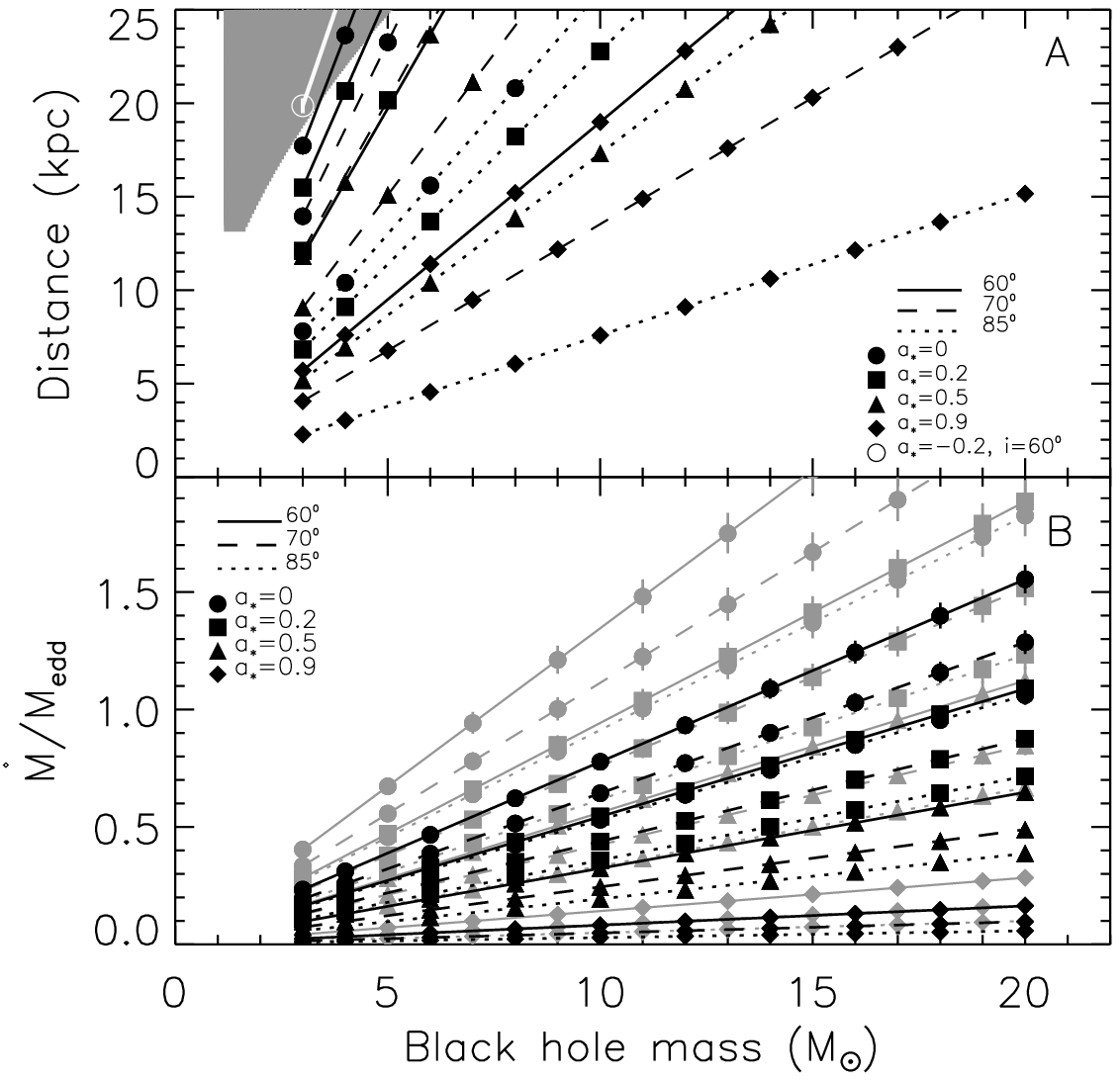} 
 \caption{Variation of distance (panel (a)) and mass accretion rate 
(panel (b)) as a function of mass for different inclination angles
and spins, as shown in the legend. For the purpose of clarity, symbols
are shown for alternate values of mass, and mass accretion rate is 
shown only for the maximum (gray) and the minimum (black) of the five values.
The gray region in the top left corner of panel (a) shows the allowed
range for the black hole mass and distance from the luminosity argument.} 
\label{mdplot}
\end{figure}

\subsection{Model: {\tt KERRBB}}
In the second part, we fitted the spectra using the model {\tt CONS*PHABS*(SIMPL$\otimes$KERRBB)}. 
Apart from mass,
distance, inclination, and spin as independent fit parameters, {\tt KERRBB}
has parameters governing the second-order effects such as `spectral hardening factor',
`returning radiation' and `limb darkening'. We, however, froze these
parameters to their default values. Typically {\tt KERRBB} is used for disk-dominated 
spectra with luminosity <0.3 $L_{Edd}$, however, \citet{steiner2} have shown that 
{\tt SIMPL}$\otimes${\tt KERRBB} can be used to accurately describe spectra 
with higher luminosities as well. 
In this case, we 
tied all {\tt KERRBB} parameters except N$_H$ and mass accretion rate across 
the five phases. Further, the normalization was frozen to 1.0. Since all 
system parameters cannot be fitted simultaneously, we selected specific
values for mass, inclination, and spin and fitted for mass accretion rate and 
distance. In order to systematically investigate the parameter space,
we varied the black hole mass from 3 $M_\odot$ to 20 $M_\odot$, inclination
angle from $50^\circ$ to $85^\circ$, and spin from 0 to 0.9. Results of 
this analysis have been shown in Figures 3 and 4. Figure 3 shows variation 
of interstellar absorption, N$_H$, two parameters of {\tt SIMPL} -- 
power-law index, $\Gamma$ and the scattering fraction, $f_{sc}$, and 
fit statistic, $\chi_{\nu}^2$ for the five phases. Since these parameters 
are more or less the same for any combination of system parameters (mass, 
distance, inclination, and spin), these are shown without any distinction.
Again it was found that neither the Fe-line nor the reflection component is
required for fitting data with the F-test chance improvement probability being
>88\% for all combinations of mass, distance, and inclination.
For a combination of mass $M$, inclination $i$ (>50$^\circ$), and spin 
$a_\ast$, the fitted values of distance $D$ (top panel) and mass accretion 
rate $\dot{M}_{Edd}$ (bottom panel) have been shown in Figure~\ref{mdplot}. 
Only maximum (phase 1) and minimum (phase 3) mass accretion rates have been 
shown in Figure~\ref{mdplot} for the sake of clarity. Error bars (90\% 
confidence limit) are smaller than the symbols for most of the points in plot.
This indicates that the distance and the accretion rate are very well 
constrained for a given combination of system parameters.

Figure~\ref{mdplot} can be used to put some significant limits on the spin 
of the black hole with some independent constraints on either black hole 
mass or distance. One such independent constraint comes from the total 
luminosity argument during `heartbeat' oscillations. \citet{neilsen1, neilsen2}
studied such oscillations in \grs{} and suggested that the radiation pressure 
instability augmented by the local Eddington limit within inner 
accretion disk as the origin of such variability pattern. However, this 
mechanism requires high accretion rate. \citet{neilsen1} showed that 
during the peak of `heartbeat', the bolometric disk luminosity is typically 
as high as 80\%--90 \% of the Eddington luminosity. Given the similarity 
between the variability patterns in \grs{} and \igr{}, it is natural to 
assume a similar mechanism operating in both the sources. In this way, the 
observed flux during the peak of the `heartbeat' oscillations can be used 
to estimate the distance for a given mass or in other words `heartbeat' can 
be used as a standard candle in order to constrain the parameters.

We found the best-fit value of the peak bolometric flux as
4.10 $\times10^{-9}$ erg s$^{-1}$cm$^{-2}$. Figure 1 shows that the peak 
flux in individual bursts is not the same and there is a burst-to-burst 
variation in the peak flux values. Therefore, we have assumed the range of 
peak flux to be (3.0 -- 5.0) $\times10^{-9}$ erg s$^{-1}$ cm$^{-2}$ and a more 
conservative range of peak luminosity to be 60\%-90\% of Eddington luminosity. 
Thus the obtained possible $M-D$ range is shown as a gray area in the top left 
corner of Figure~\ref{mdplot}. It can be seen that the points within this 
region correspond to inclination angle <60$^\circ$ and spin <0.2. However, 
the lower limit on inclination was found to be $\sim$76$^\circ$ from 
{\tt DISKPN} normalization. This presents a tantalizing possibility of black 
hole spin to be retrograde indicating a black hole spin in opposite direction 
to the accretion disk. Further, we have found a lower limit of $\sim$53$^\circ$
on inclination angle from 90\% upper confidence limit of {\tt DISKPN} 
normalization. The points corresponding to $i\geq$50$^\circ$ require spin 
to be <0.2. Thus we exclude the possibility of high spin and it appears 
that the main reason for the observed faintness of \igr{} is very low (or 
even negative) spin of the black hole, in contrast to \grs{}, which 
is known to have a very high spin \citep{mc}.

It can be seen that even for the exotic type of black hole with mass <3 
$M_\odot$ \citep{altamirano2}, this inference on black hole spin remains 
valid. Further, in order to verify the effect of other frozen parameters 
of {\tt KERRBB}, such as returning radiation, limb darkening, and inner boundary 
stress, we have also enabled all of them (one by one as well as all together) 
and found that the lines corresponding to a particular combination of spin 
and inclination angle move away from the shaded region. We verified that
the same is valid when {\tt POWERLAW} is used instead of {\tt SIMPL}. Thus 
for any combination of these parameters or model for the high-energy tail, 
the black hole spin is required to be either very low or negative.

The 66 Hz HFQPO of \igr{} detected by \citet {altamirano3} may also be 
considered as an independent constraint on black hole mass,  if it is assumed 
to be related to mass. In this case, the black hole mass in \igr{} has to 
be $\sim15 M_{\odot}$. For such a high black hole mass, Figure 4(a) does not 
provide any constraint on the distance, however, both the inclination angle 
and the spin are required to be >70$^\circ$ and >0.7, respectively. For these 
high values of inclination angle and spin, Figure 4(b) shows that the required 
accretion rate is only a small fraction of the Eddington accretion rate. 
Thus, if we assume that the 66 Hz QPO is related to the black hole mass, the 
basic accretion process giving rise to the apparent similar variability of 
\grs{} and \igr{} has to be different.
However, it is more likely, as also suggested by \citet{altamirano3}, 
that the 66 Hz HFQPO is not directly related to the black hole
mass. In that case, the previous argument for a low black hole mass
based on the apparent flux and similar accretion process between 
\igr{} and \grs{} holds, and the inference of low or negative spin remains
valid. Overall this work indicates that the black hole spin may not play 
a significant role in generating the observed extreme variability and such 
a behavior is generated mainly by a high accretion rate.

\begin{center}
\begin{table*}
\centering
 \caption{Parameters Obtained from a Simultaneous Fit to the Spectra of Five Phases with Model \tt {CONS*PHABS*(SIMPL$\otimes$DISKPN)}}
    \begin{tabular}{p{3.5cm} c c c c c}
    \hline
    \hline
      & Phase 1 & Phase 2 & Phase 3 & Phase 4 & Phase 5\\
\hline
$N_H$ ($\times$10$^{22}$cm$^{-2}$) & 0.93$^{+0.03}_{-0.03}$ & 0.84$^{+0.02}_{-0.02}$ & 0.81$^{+0.01}_{-0.02}$ & 0.81$^{+0.01}_{-0.01}$ & 0.88$^{+0.01}_{-0.01}$\\

T$_{max}$ (keV) & 1.24$^{+0.05}_{-0.05}$ & 1.11$^{+0.01}_{-0.01}$ & 1.02$^{+0.01}_{-0.01}$ & 1.05$^{+0.01}_{-0.01}$ & 1.11$^{+0.01}_{-0.01}$ \\

R$_{in}$ (R$_g$) & 24.8$^{+2.9}_{-2.4}$ & 24.9$^{+1.2}_{-1.1}$ & 28.2$^{+1.3}_{-1.2}$ & 26.3$^{+1.0}_{-0.9}$ & 27.3$^{+1.0}_{-1.0}$\\
$\Gamma$ & 3.57$^{+0.25}_{-0.28}$ & 2.66$^{+0.10}_{-0.10}$ & 2.56$^{+0.06}_{-0.06}$ & 2.52$^{+0.04}_{-0.04}$ & 2.70$^{+0.06}_{-0.06}$\\
Scattered fraction & 0.47$^{+0.02}_{-0.28}$ & 0.27$^{+0.02}_{-0.02}$ & 0.36$^{+0.02}_{-0.02}$ & 0.37$^{+0.01}_{-0.01}$ & 0.35$^{+0.02}_{-0.02}$\\
Disk flux$^a$ & 2.10 & 1.23 & 1.01 & 1.06 & 1.47\\
Total flux$^a$ & 2.54 & 1.53 & 1.42 & 1.49 & 1.92\\
$\tt DISKPN$ norm$^b$ & \multicolumn{4}{c}{4.0$^{+6.4}_{-3.7}\times10^{-4}$}\\
\hline
\multicolumn{6}{l}{$a$: Unabsorbed flux in units of $\times$10$^{-9}$erg s$^{-1}$cm$^{-2}$ for 2 -- 10 keV energy range.}\\
\multicolumn{6}{l}{~~~ Unabsorbed bolometric disk flux for phase 1 is 4.10$\times$10$^{-9}$erg s$^{-1}$cm$^{-2}$.}\\
\multicolumn{6}{l}{~~~ (See the text for discussion).}\\

\multicolumn{6}{l}{$b$: Since spectral fitting could not constrain the lower limit of norm, it was}\\
\multicolumn{6}{l}{~~~calculated from the extreme values of $M$, $D$ and $i$.} \\

    \end{tabular} 
 \end{table*}

\end{center}

\section{Conclusions}
\grs{} was so far the only BHB exhibiting extreme variability and 
spectral changes over timescales as short as a few tens of seconds. 
Observations of similar variability in another BHB \igr{}
establishes that such extreme variability of \grs{} is not due to 
some specific coincidences unique to one particular BHB, but is a more
generic phenomenon. 
Here, we presented results of simultaneous
fitting of 0.7--35.0 keV spectra during different phases of the
`heartbeat' oscillations in \igr{}, which indicate that the most likely 
difference between \grs{} and \igr{} is in the spins of the respective 
black holes. While the black hole in \grs{} is known to be rotating with
high spin, the black hole in \igr{} is found to have a very low spin. In
fact, for inclination $\sim$70$^\circ$, which is favored by the data,
the black hole could very well have a retrograde spin. In this case,
\igr{} would be the first known astrophysical source having a retrograde spin.
Even though theoretically possible, such a  scenario would be very challenging 
to explain from the point of view of evolution of such a system.

\acknowledgments
This research has made use of data obtained from High Energy 
Astrophysics Science Archive Research Center (HEASARC), provided by 
NASA's Goddard Space Flight Center. We thank A. R. Rao for useful
discussions. We also thank the anonymous referee for very useful comments.

\bibliographystyle{apj}

\begin{thebibliography}{27}
\expandafter\ifx\csname natexlab\endcsname\relax\def\natexlab#1{#1}\fi

\bibitem[{{Altamirano} \& {Belloni}(2012)}]{altamirano3}
{Altamirano}, D., \& {Belloni}, T. 2012, \apjl, 747, L4

\bibitem[{{Altamirano} {et~al.}(2012){Altamirano}, {Wijnands}, {Belloni}, \&
  {Motta}}]{altamirano1}
{Altamirano}, D., {Wijnands}, R., {Belloni}, T., \& {Motta}, S. 2012, The
  Astronomer's Telegram, 3913, 1

\bibitem[{{Altamirano} {et~al.}(2011){Altamirano}, {Belloni}, {Linares}, {van
  der Klis}, {Wijnands}, {Curran}, {Kalamkar}, {Stiele}, {Motta},
  {Mu{\~n}oz-Darias}, {Casella}, \& {Krimm}}]{altamirano2}
{Altamirano}, D., {Belloni}, T., {Linares}, M., {et~al.} 2011, \apjl, 742, L17

\bibitem[{{Belloni} {et~al.}(2000){Belloni}, {Klein-Wolt}, {M{\'e}ndez}, {van
  der Klis}, \& {van Paradijs}}]{belloni}
{Belloni}, T., {Klein-Wolt}, M., {M{\'e}ndez}, M., {van der Klis}, M., \& {van
  Paradijs}, J. 2000, \aap, 355, 271

\bibitem[{{Capitanio} {et~al.}(2012){Capitanio}, {Del Santo}, {Bozzo},
  {Ferrigno}, {De Cesare}, \& {Paizis}}]{capitanio1}
{Capitanio}, F., {Del Santo}, M., {Bozzo}, E., {et~al.} 2012, \mnras, 422, 3130

\bibitem[{{Capitanio} {et~al.}(2006){Capitanio}, {Bazzano}, {Ubertini},
  {Zdziarski}, {Bird}, {De Cesare}, {Dean}, {Stephen}, \&
  {Tarana}}]{capitanio2}
{Capitanio}, F., {Bazzano}, A., {Ubertini}, P., {et~al.} 2006, \apj, 643, 376

\bibitem[{{Capitanio} {et~al.}(2009){Capitanio}, {Giroletti}, {Molina},
  {Bazzano}, {Tarana}, {Kennea}, {Dean}, {Hill}, {Tavani}, \&
  {Ubertini}}]{capitanio3}
{Capitanio}, F., {Giroletti}, M., {Molina}, M., {et~al.} 2009, \apj, 690, 1621

\bibitem[{{Done} {et~al.}(2004){Done}, {Wardzi{\'n}ski}, \&
  {Gierli{\'n}ski}}]{done}
{Done}, C., {Wardzi{\'n}ski}, G., \& {Gierli{\'n}ski}, M. 2004, \mnras, 349,
  393

\bibitem[{{Fender} \& {Belloni}(2004)}]{fenbel}
{Fender}, R., \& {Belloni}, T. 2004, \araa, 42, 317

\bibitem[{{Fender} {et~al.}(1999){Fender}, {Garrington}, {McKay}, {Muxlow},
  {Pooley}, {Spencer}, {Stirling}, \& {Waltman}}]{fender}
{Fender}, R.~P., {Garrington}, S.~T., {McKay}, D.~J., {et~al.} 1999, \mnras,
  304, 865

\bibitem[{{Fryer} \& {Kalogera}(2001)}]{fryer}
{Fryer}, C.~L., \& {Kalogera}, V. 2001, \apj, 554, 548

\bibitem[{{Gierli{\'n}ski} {et~al.}(1999){Gierli{\'n}ski}, {Zdziarski},
  {Poutanen}, {Coppi}, {Ebisawa}, \& {Johnson}}]{gierlinski}
{Gierli{\'n}ski}, M., {Zdziarski}, A.~A., {Poutanen}, J., {et~al.} 1999,
  \mnras, 309, 496

\bibitem[{{King} {et~al.}(2012){King}, {Miller}, {Raymond}, {Fabian},
  {Reynolds}, {Kallman}, {Maitra}, {Cackett}, \& {Rupen}}]{king}
{King}, A.~L., {Miller}, J.~M., {Raymond}, J., {et~al.} 2012, \apjl, 746, L20

\bibitem[{{Krimm} \& {Kennea}(2011)}]{krimmkennea}
{Krimm}, H.~A., \& {Kennea}, J.~A. 2011, The Astronomer's Telegram, 3148, 1

\bibitem[{{Krimm} {et~al.}(2011){Krimm}, {Barthelmy}, {Baumgartner},
  {Cummings}, {Fenimore}, {Gehrels}, {Kennea}, {Markwardt}, {Palmer},
  {Sakamoto}, {Skinner}, {Stamatikos}, {Tueller}, \& {Ukwatta}}]{krimm}
{Krimm}, H.~A., {Barthelmy}, S.~D., {Baumgartner}, W., {et~al.} 2011, The
  Astronomer's Telegram, 3144, 1

\bibitem[{{Kuulkers} {et~al.}(2003){Kuulkers}, {Lutovinov}, {Parmar},
  {Capitanio}, {Mowlavi}, \& {Hermsen}}]{kuulkers}
{Kuulkers}, E., {Lutovinov}, A., {Parmar}, A., {et~al.} 2003, The Astronomer's
  Telegram, 149, 1

\bibitem[{{Li} {et~al.}(2005){Li}, {Zimmerman}, {Narayan}, \&
  {McClintock}}]{li}
{Li}, L.-X., {Zimmerman}, E.~R., {Narayan}, R., \& {McClintock}, J.~E. 2005,
  \apjs, 157, 335

\bibitem[{{McClintock} {et~al.}(2006){McClintock}, {Shafee}, {Narayan},
  {Remillard}, {Davis}, \& {Li}}]{mc}
{McClintock}, J.~E., {Shafee}, R., {Narayan}, R., {et~al.} 2006, \apj, 652, 518

\bibitem[{{Mirabel} {et~al.}(1998){Mirabel}, {Dhawan}, {Chaty}, {Rodriguez},
  {Marti}, {Robinson}, {Swank}, \& {Geballe}}]{mirabel}
{Mirabel}, I.~F., {Dhawan}, V., {Chaty}, S., {et~al.} 1998, \aap, 330, L9

\bibitem[{{Neilsen} {et~al.}(2011){Neilsen}, {Remillard}, \& {Lee}}]{neilsen1}
{Neilsen}, J., {Remillard}, R.~A., \& {Lee}, J.~C. 2011, \apj, 737, 69

\bibitem[{{Neilsen} {et~al.}(2012){Neilsen}, {Remillard}, \& {Lee}}]{neilsen2}
---. 2012, \apj, 750, 71

\bibitem[{{Pahari} {et~al.}(2011){Pahari}, {Yadav}, \&
  {Bhattacharyya}}]{pahari}
{Pahari}, M., {Yadav}, J., \& {Bhattacharyya}, S. 2011, ArXiv e-prints

\bibitem[{{Revnivtsev} {et~al.}(2003){Revnivtsev}, {Gilfanov}, {Churazov}, \&
  {Sunyaev}}]{revnivtsev}
{Revnivtsev}, M., {Gilfanov}, M., {Churazov}, E., \& {Sunyaev}, R. 2003, The
  Astronomer's Telegram, 150, 1

\bibitem[{{Rodriguez} {et~al.}(2011){Rodriguez}, {Corbel}, {Caballero},
  {Tomsick}, {Tzioumis}, {Paizis}, {Cadolle Bel}, \& {Kuulkers}}]{rodriguez}
{Rodriguez}, J., {Corbel}, S., {Caballero}, I., {et~al.} 2011, \aap, 533, L4

\bibitem[{{Steiner} {et~al.}(2009{\natexlab{a}}){Steiner}, {McClintock},
  {Remillard}, {Narayan}, \& {Gou}}]{steiner2}
{Steiner}, J.~F., {McClintock}, J.~E., {Remillard}, R.~A., {Narayan}, R., \&
  {Gou}, L. 2009{\natexlab{a}}, \apjl, 701, L83

\bibitem[{{Steiner} {et~al.}(2009{\natexlab{b}}){Steiner}, {Narayan},
  {McClintock}, \& {Ebisawa}}]{steiner}
{Steiner}, J.~F., {Narayan}, R., {McClintock}, J.~E., \& {Ebisawa}, K.
  2009{\natexlab{b}}, \pasp, 121, 1279

\bibitem[{{Vadawale} {et~al.}(2003){Vadawale}, {Rao}, {Naik}, {Yadav},
  {Ishwara-Chandra}, {Pramesh Rao}, \& {Pooley}}]{vadawale}
{Vadawale}, S.~V., {Rao}, A.~R., {Naik}, S., {et~al.} 2003, \apj, 597, 1023

\end{thebibliography}

\end{document}